\documentclass[pdflatex,sn-mathphys]{sn-jnl}% Math and Physical Sciences Reference Style
\usepackage[english]{babel}
\usepackage{amsmath}
\usepackage{amsfonts}
\usepackage{empheq}
\usepackage{stmaryrd}
\usepackage{physics}
\usepackage{mathtools}
\usepackage{caption}
\usepackage{subcaption}
\usepackage[export]{adjustbox}
\usepackage{yfonts}
\usepackage{mathrsfs}
\usepackage{url}
\usepackage{stmaryrd, physics, mathtools}
\usepackage{hyperref}
\usepackage{doi}

\usepackage{multirow}
\usepackage{rotating}
\usepackage{graphics}
\usepackage{titlesec}
\jyear{2021}%

\theoremstyle{thmstyleone}%
\theoremstyle{thmstyletwo}%

\theoremstyle{thmstylethree}%

\begin{document}

\title[Application of a dispersive wave HSM model in Ria Formosa]{Application of a dispersive wave hydro-sediment-morphodynamic model in the Ria Formosa lagoon}


\author*[1]{\fnm{Kazbek} \sur{Kazhyken}}\email{kazbek.kazhyken@gmail.com}

\author[1,3,4]{\fnm{Eirik} \sur{Valseth}}\email{eirik@oden.utexas.edu}

\author[2]{\fnm{Juha} \sur{Videman}}\email{jvideman@math.tecnico.ulisboa.pt}

\author[1,2]{\fnm{Clint} \sur{Dawson}}\email{clint@oden.utexas.edu}

\affil[1]{\orgdiv{Oden Institute for Computational Engineering and Sciences}, \orgname{The University of Texas at Austin}, \orgaddress{\city{Austin}, \postcode{78712}, \state{TX}, \country{USA}}}

\affil[2]{\orgdiv{CAMGSD/Departamento de Matem\'atica}, \orgname{Universidade de Lisboa}, \orgaddress{\city{Lisbon}, \postcode{1049-001}, \country{Portugal}}}

\affil[3]{\orgdiv{Department of Data Science}, \orgname{The Norwegian University of Life Sciences}, \orgaddress{\city{Ås}, \postcode{1433}, \country{Norway}}}

\affil[4]{\orgdiv{Department of Numerical Analysis and Scientific Computing},\orgname{Simula Research Laboratory}, \orgaddress{\city{Oslo}, \postcode{0164}, \country{Norway}}}


\abstract{Results of an application of a dispersive wave hydro-sediment-morphodynamic model in the western circulation cell of the Ria Formosa lagoon located in the Algarve region of the southern Portugal are presented. This area of interest has a couple of features that complicate the application of the dispersive wave model: (1) the area has a complex irregular geometry with a number of barrier islands that separate the lagoon from the Atlantic Ocean, artificial and naturally occurring tidal inlets, and a number of curling channels inside the lagoon that interconnect the inlets and serve as waterways between the lagoon settlements; (2) the tidal range in the area can reach up to 3.5 m; therefore, the terrain inside the lagoon is characterized by vast salt marshes and tidal flats, and the wetting-drying process is a key component of any hydrodynamic simulation in this area. A model representation of the area has been developed by generating an unstructured finite element mesh of the circulation cell, and collecting data on parameters that characterize the tidal waves in the area, and bottom friction and sediment transport models used in the simulations. The results of the simulations indicate that the dispersive wave model can be applied in coastal areas with nontrivial underlying physical processes, and complex irregular geometries. Moreover, the dispersive term of the model is capable of capturing additional flow characteristics that are otherwise not present in hydrodynamic simulations that involve the nonlinear shallow water equations; and these additional flow features can, in their turn, affect the resulting sediment transport and bed morphodynamic process simulations.}

\keywords{Ria Formosa, dispersive waves, sediment transport, discontinuous Galerkin methods}

\pacs[Statements and Declarations]{The authors have no competing interests to declare that are relevant to the content of this article.}

\pacs[MSC Classification]{65M60, 76B15, 86-08, 86-10}

\maketitle

\section{Introduction}
Coastal areas have continuously attracted humans as a suitable habitat due to a number of reasons, such as their rich supply of resources that could be utilized for sustenance, their facility as an access point for marine transportation which is an important logistical component of the world trade and commerce, and their potential as a location for various cultural and recreational activities \cite{neuman_etal_2015}. According to the Census data, the population of the United States that resides in coastal areas including areas around the Great Lakes has grown from nearly 105 million in 2000 to a little over 123 million in 2010, or in relative terms from 37\% to 39\% of the total U.S. population \cite{crowell_etal_2007, ache_et_al_2013}. Worldwide the scale of the population living in areas that are no more than 100 km away from the coast is comparable and reaches 40\% of the total world population that compounds to a little less than 3 billion people \cite{un}. Therefore, it is of no coincidence that studying coastal environments, including the effects of tides and currents on coastal morphology, continues to have a clear engineering and scientific relevance; and any substantial study of a coastal engineering problem, for instance coastline and beach erosion prevention, natural hazard protection, building and maintenance of coastal structures, is heavily reliant on the mathematical modeling of hydrodynamics, sediment transport, and bed morphodynamics in coastal areas.

Coastal morphodynamics is a complex physical process that is driven by complicated non-linear interactions between water waves forced by astronomical tides, winds, and long-wave currents, sediment transport, and bed morphodynamics. In addition to these naturally occurring forces, anthropogenic deformations of coastal areas have the potential to magnify the interactions and accelerate changes in coastal morphology. Changes in coastal morphology propelled by natural and anthropogenic forces are sufficient to negatively affect coastal infrastructure and environment. For instance, excessive erosion of seabed due to scouring may have a negative impact on structural integrity of piers, levees and other coastal structures. Excessive sediment deposition in a harbor may damage its walls, and negatively impact its economic viability by increasing the harbor's operating costs due to required dredging; therefore, its effects play an important role in harbor planning and construction. Among environmental concerns are degradation of natural habitats of endangered and protected species due to shoreline and beach erosion, and the effects of sediment transport on contaminant deposition and distribution where sediment deposits may serve as sinks or sources for dangerous contaminants depending on surrounding physico-chemical conditions.

A chain of hydrodynamic, sediment transport, and bed morphodynamic processes forms a closed-loop system where sediment transport is affected by flow parameters, such as flow velocity and turbulence, bed morphology changes as a consequence of sediment transport processes where sediment is net eroded from one area to be net deposited into another area, and, finally, the loop is closed when the flow parameters are affected by changes in the bathymetry due to the effects of sediment entrainment and deposition. This physical system of the interrelated processes is referred to as a hydro-sediment-morphodynamic process. A mathematical model designed to describe the process, by the nature of this process, is formed by tying together non-linear hydrodynamic, sediment transport, and bed morphodynamic models along with modeling their mutual interactions. A number of models, ranging from one to three dimensions, suitable for an engineering analysis of a coastal hydro-sediment-morphodynamic process have been developed. One such model is formed by the shallow water hydro-sediment-morphodynamic (SHSM) equations (see e.g. Cao \emph{et al.} \cite{cao_etal_2002}), which are derived by integrating and averaging three-dimensional mass and momentum conservation equations of motion (see e.g. see Wu \cite{wu_2007}). In the SHSM equations, the hydrodynamic model is represented by the nonlinear shallow water equations that provide sufficiently accurate approximation of hydrodynamics if the shallowness parameter that is defined as the ratio between the depth and length scales of the flow is less than unity. However, this hydrodynamic model does not have a capacity to capture wave dispersion effects; and, therefore, an application of the SHSM equations is not feasible in areas where the dispersion effects are prevalent. 

An alternative dispersive wave hydro-sediment-morphodynamic model that employs the Green-Naghdi equations as its hydrodynamic model is developed in \cite{kazhyken_etal_2021_0, kazhyken_etal_2021_1}, and can be used in applications where the dispersive effects play a significant role in hydrodynamics. The capacity to capture dispersive wave effects in the developed model comes, however, with a greater analytical and numerical complexity. The major point of complexity is introduced through the dispersive term of the model; this term models effects of wave dispersion on flow velocities and is an integral part of the Green-Naghdi equations. Therefore, a Strang operator splitting technique from \cite{strang_1968} has been employed to treat the developed model numerically. The model has been split into two parts: (1) the dispersive correction part comprised of the dispersive term only, (2) the remaining shallow water hydro-sediment-morphodynamic equations. The hybridized discontinuous Galerkin developed by Samii and Dawson in \cite{samii_and_dawson_2018} method has been employed to develop a numerical solution algorithm for the dispersive correction part; and for the remaining SHSM part of the model hybridized and regular discontinuous Galerkin methods have been employed to develop the numerical algorithms \cite{kazhyken_etal_2021_1}. The  numerical algorithms were implemented in a C++ software package \footnote{The software is under development on the date of the publication, and can be accessed at \url{www.github.com/UT-CHG/dgswemv2}. Should there be any questions, comments, or suggestions, please contact the developers through the repository issues page.} and have been validated against a number of simulation benchmarks in \cite{kazhyken_etal_2021_0, kazhyken_etal_2021_1}.

The purpose of the presented work is to simulate hydro-sediment-morphodynamic processes in the Ria Formosa lagoon in Portugal with the developed dispersive wave model. The lagoon stretches about 55km along the southern coast of Portugal; it is separated from the Atlantic Ocean by a series of barrier islands, and has five tidal inlets. The lagoon presents itself as a fairly complicated physical system; therefore, applying the developed model in this coastal area proves to be a challenging problem. Numerous works have been focused on modeling hydro-sediment-morphodynamic phenomena in the Ria Formosa lagoon. Pacheco \emph{et al.} present a long-term sediment budget computations for tidal inlets of the lagoon in \cite{pacheco_etal_2008}. Simulations of a short-term storm driven morphological evolution of the Anc\~ao inlet of the lagoon have been studied by Williams and Pan in \cite{williams_pan_2011}. Attempts to estimate parameters for empirical sediment transport and bed morphodynamic models have been done by Pacheco \emph{et al.} in \cite{pacheco_etal_2011}. A number of numerical simulations focused on the lagoon have been carried out using the well-known Delft3D software package, e.g. Gonz\'ales-Gorben\~a \emph{et al.} in \cite{gorbena_etal_2017}, or Carrasco \emph{et al.} in \cite{carrasco_etal_2018}. These simulations employed depth averaged hydrodynamic models; among other simulations that use a depth-averaged hydrodynamic model are works by Dias \emph{et al.} in \cite{dias_etal_2009}, or Salles \emph{et al.} in \cite{salle_etal_2005}. In the INDIA project, see, e.g., \cite{do2002sudden,williams2003tidal}, numerical models of dispersive wave hydrodynamics are considered in the Ria Formosa lagoon. The models developed are based on numerical approximations using both finite element and finite difference methods.
To the best of our knowledge there is no work that attempts to model hydrodynamics in the lagoon using a dispersive wave model based on hybridized discontinuous Galerkin discretizations. We attempt to simulate hydro-sediment-morphodynamic processes in the lagoon to test the capabilities of the developed model to carry out such simulations in irregularly shaped domains with non-trivial underlying physical processes. 

The rest of this work is organized as follows. Section 2 provides a brief description of the lagoon and its characteristics. In Section 3, the governing equations for simulations are presented. Section 4 provides details on the work that has been accomplished to develop a finite element mesh of the western circulation cell of the lagoon. In Section 5, results of the simulations of hydro-sediment-morphodynamic processes in the Ria Formosa lagoon are presented. Final conclusions are given in Section 6.

\section{Area of interest}
\begin{figure}[h!]
\centering
\includegraphics[width=4in]{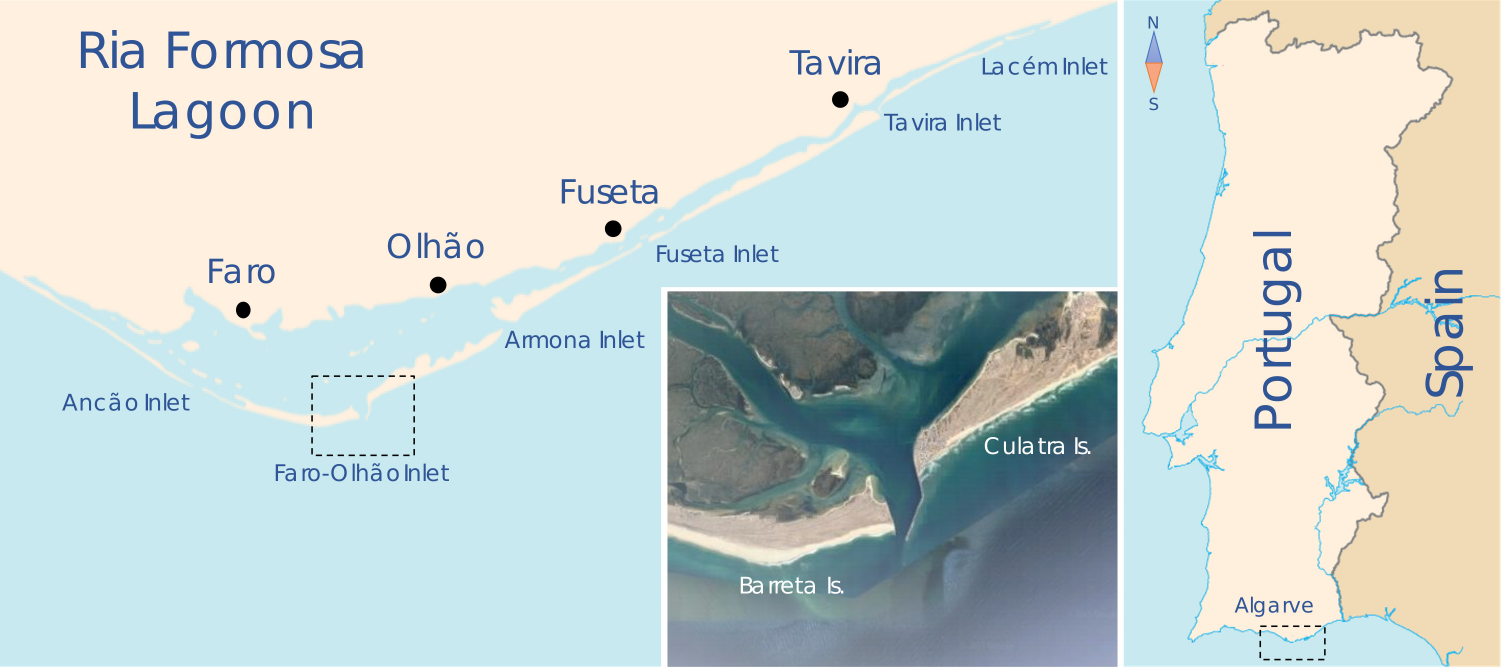}
\caption{The Ria Formosa lagoon diagram. Reproduced from the same data as in~\cite{kazhyken_etal_2021_0}.}
\label{Fig:Ria}
\end{figure}

The Algarve is the southernmost region of Portugal, with its capital in the city of Faro, that has a long tradition of using its coast as the main provider for its local economy. The area is one of the most important fishing regions in Portugal; and it has recently transformed into one of the major tourist attractions of the Iberian peninsula due to its coastal area characterized by offshore sandbars, and unique wetland ecosystems such as the Ria Formosa lagoon \cite{noronha_vaz_etal_2013}. The lagoon has a complex cuspate-shaped 55 km long (east to west) and 6 km wide (north to south at its widest part) geometry \cite{carrasco_etal_2018}, and is separated from the Atlantic Ocean with two peninsulas, the Anc\~ao and Cacela at its western and eastern ends, respectively, and five barrier islands, the Barreta, Culatra, Armona, Tavira and Cabanas islands spanning west to east. The lagoon is connected to the Atlantic Ocean with six naturally occurring and artificial inlets: the Anc\~ao, Faro-Olh\~ao, Armona, Fuzeta, Tavira and Lac\`em inlets, which are responsible for the water, sediment, chemical, and nutrient transport between the Atlantic Ocean and the lagoon marshes \cite{ceia_etal_2010}. Refer to Fig.~\ref{Fig:Ria} for an overview of the locations on the southern coast of Portugal. Nearly 75\% of the water in the lagoon is exchanged daily due to the tides, with the majority of the exchange occurring through the Faro-Olh\~ao inlet. The currents induced by the tides in the inlets are strong, and model results from~\cite{gorbena_etal_2017} show (depth-averaged) tidal velocities of more than 3$\text{ms}^{-1}$ during spring-tide ebb.
% and can reach 2.2 $\text{ms}^{-1}$ and 1.6 $\text{ms}^{-1}$ at the Faro-Olh\~ao inlet during the ebb and flood, respectively \cite{salles_etal_2005}.
% \cite{gorbena_etal_2017}
%
%

The lagoon is characterized by vast salt marshes, tidal flats, and an intricate network of naturally occurring and partially dredged channels. The wetted area of the lagoon, which includes a large intertidal zone, reaches up to 105 $\text{km}^2$; and, according to the Hydrographic Institute of Portugal, the submerged area of the lagoon is 53 $\text{km}^2$ at high water and between 14 and 22 $\text{km}^2$ at low water \cite{costa1986mares}. The salt marshes primarily contain silt and fine sand, and are penetrated with a high density of meandering tidal creeks \cite{dias_etal_2009}. The tidal flats and salt marshes cover more than 60\% of the total area of the lagoon \cite{carrasco_etal_2018}. The average depth of the partially dredged navigable channels is 6 m, although in most areas the channel depth reaches 2 m only \cite{salles_etal_2005}. The system lacks any significant river. Five small rivers and 14 streams, most of which go completely dry during the summer, flow into the Ria Formosa lagoon; therefore, freshwater discharges are negligible relative to tidal prisms and baroclinic effects are minor \cite{dias_etal_2009}. Salinity values in the lagoon are usually close to those observed
in the adjacent coastal waters \cite{soares_etal_2020}.

\begin{figure}[h!]
\centering
\includegraphics[width=4in]{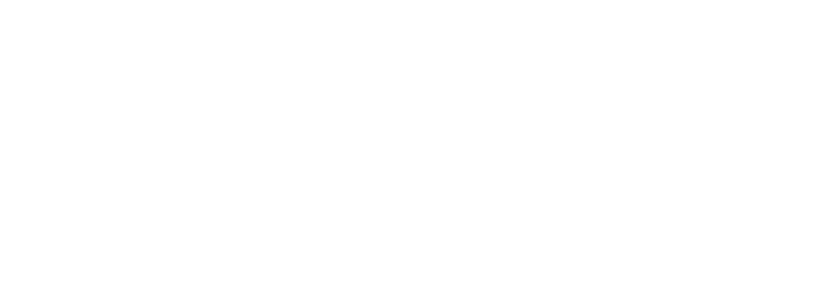}
\caption{The tidal inlets of the western hydrodynamic circulation cell of the Ria Formosa lagoon. Reproduced from the same data as in~\cite{kazhyken_etal_2021_0}.}
\label{Fig:RM_p1}
\end{figure}

Based on the interconnections of the inlets with the channels, the lagoon is divided into three relatively independent hydrodynamic circulation cells: (1) the western cell served by the Anc\~ao, Faro-Olh\~ao, and Armona inlets (cf. Fig. \ref{Fig:RM_p1}), (2) the central cell served by the Fuseta inlet, and (3) the eastern cell served by the Tavira and Cacela inlets \cite{silva_etal_2014}.
The western cell is the largest and has complex hydrodynamic characteristics; moreover, it has been a subject of a number of anthropogenic interventions. The Anc\~ao, Faro-Olh\~ao, and Armona inlets capture nearly 90\% of the total tidal prism of the lagoon: during the spring tides the inlets capture 61\%,  23\%, and 8\% of the total flow, respectively, and during the neap tides the Faro-Olh\~ao and Armona inlets capture 45\% and 40\% of the flow, respectively, while an insignificant amount is contributed by the Anc\~ao inlet \cite{carrasco_etal_2018}. The three inlets present an ebb dominated behavior with higher mean ebb velocities and shorter ebb durations \cite{pacheco_etal_2007}. According to the tidal prism measurements performed between 2006 and 2007 for each inlet of the lagoon, there exists a clear circulation pattern between the Faro-Olh\~ao and Armona inlets \cite{gorbena_etal_2017}.

The tides in the area are semi-diurnal and mesotidal with the mean range of 2 m that can reach up to 3.5 m during the spring tides \cite{ferreira_etal_2016}. The wave climate in the area is moderate with the average annual significant offshore wave height of 1 m and average peak period of 8.2 s, although winter storms can produce waves with the significant height larger than 3 m \cite{costa_etal_2001}. The lagoon's cuspate shape produces two areas that are exposed to two different wave actions. The western side of the lagoon is impacted by waves originating from the west-southwest direction and produced by the North Atlantic Ocean. These waves are dominant in the area and occur nearly 68\% of time \cite{costa_etal_2001}. The longer eastern side of the lagoon is exposed to the less energetic Levante events originating from the east-southeast direction. These events account for 29\% of the wave climate \cite{costa_etal_2001}. Storms impacting the western side have higher energy and typically occur between October and March, while the ones on the eastern side primarily originate from the eastern part of the Gulf of Cadiz and occur between December and March \cite{oliveira_etal_2018}. The storm surge events in this area are relatively insignificant due to its narrow continental shelf. During extreme storm conditions the surge levels can potentially reach values close to 0.6 m \cite{gorbena_etal_2017}.

The highly dynamic morphological nature of the Ria Formosa lagoon is characterized by the migration of the tidal inlets and growth of the barrier islands. Moreover, additional hydro-sediment-morphodynamic processes, such as  longshore drift and barrier island overwash, have contributed significantly to the shape of the system \cite{gorbena_etal_2017}.  Overall, the evolution of the shoreline can be characterized by a gradual transition from coastal retreat in the west to advance in the east \cite{kombiadou_etal_2019}. The system is very sensitive to construction of coastal structures, for instance, the artificial stabilization of the Faro-Olh\~ao and Tavira inlets has let to an increased rate of coastal erosion \cite{ceia_etal_2010}. The area is also noted for its overwash susceptibility due to a storm induced coastal erosion and the sea level rise. In some instances, traces of human activity, such as foot and tractor paths, lead to formation of overwash passes which made the overwash risks even higher \cite{ceia_etal_2010}. Moreover, some of the overwashes have been linked to a growing number of human settlements, in particular the Praia de Faro and the Ilha do Farol settlements, that lead to lower dune crests and a higher risk of the overwash \cite{achab_etal_2014}.  The morphological changes can have a significant negative impact on socio-economical activities in the region,  and pose risks of destruction of its precious ecosystems \cite{noronha_vaz_etal_2013}. Therefore, monitoring and forecasting hydro-sediment-morphodynamic processes in the Ria Formosa lagoon is an essential part of understanding these morphological changes that can lead to a more effective preservation and planning methods in the process of a sustainable regional development.

\section{Governing equations}

The governing model for hydro-sediment-morphodynamic transport was developed in \cite{kazhyken_etal_2021_1} is employed here. For the sake of brevity, we do not include all details of the model herein but refer interested readers to our past work in \cite{kazhyken_etal_2021_1}.

\begin{figure}[h!]
\center
\includegraphics[width=3in]{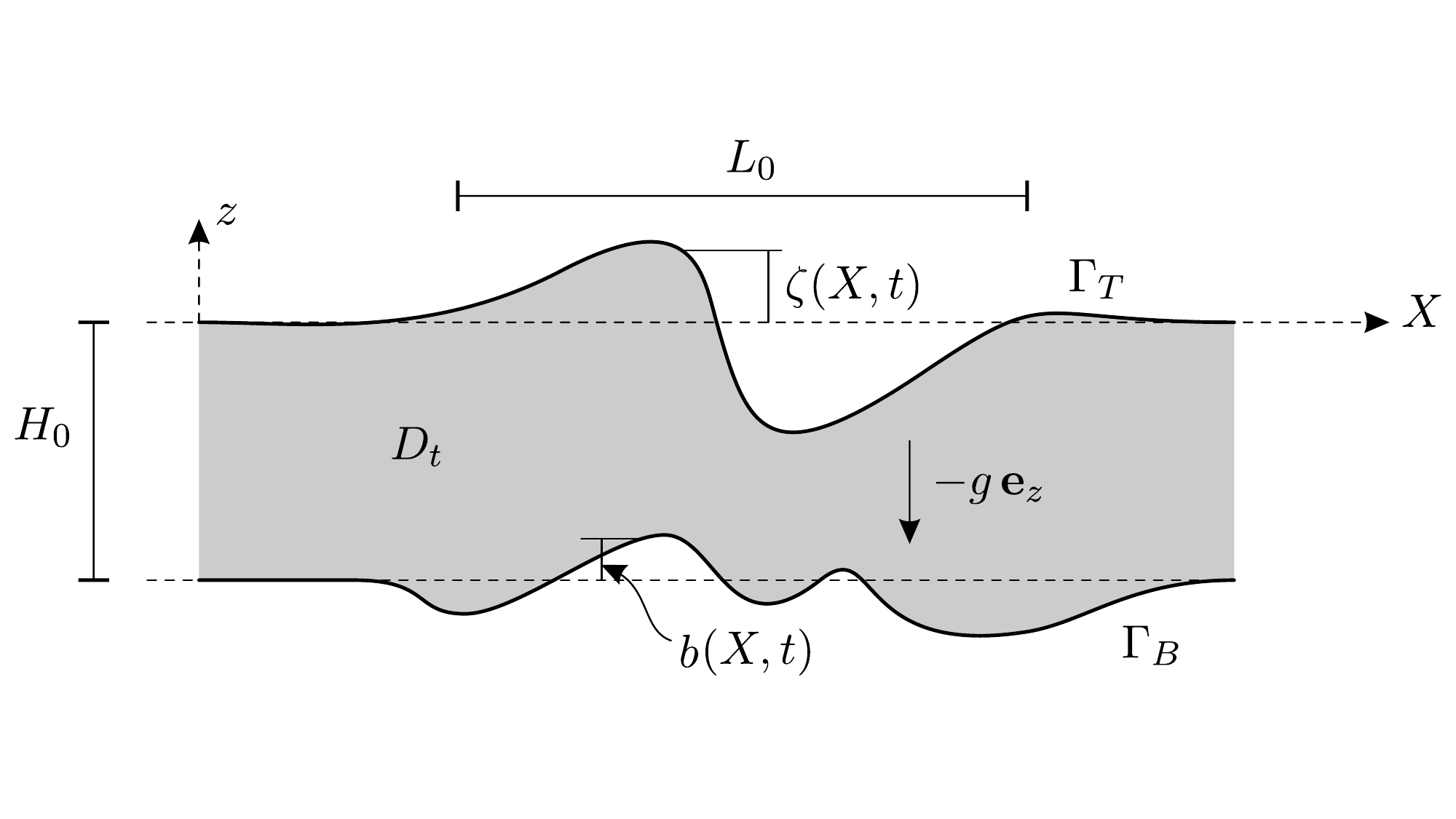}
\caption{A model representation of a body of water as a domain $D_t \subset \mathbb R^{d+1}$, figure reproduced from~\cite{kazhyken_etal_2021_0, kazhyken_etal_2021_1,samii_etal_2019,samii_and_dawson_2018}.}
\label{Fig:Domain}
\end{figure}

The model is defined on a domain $D_t \subset \mathbb R^{d+1}$ filled with a water-sediment mixture, i.e., an incompressible inviscid fluid, see Fig.~\ref{Fig:Domain}, where $d$ is the horizontal spatial dimension $(d\ge 1)$, $t$  the time variable, $\Gamma_T$ and $\Gamma_B$ are the top and bottom boundaries of the domain, respectively, $L_0$ is the characteristic length, and $H_0$ is the reference depth. We assume $\Gamma_T$ and $\Gamma_B$ can be represented as graphs, and fluid particles cannot cross these boundaries. Both boundaries vary with time, $\Gamma_B$  due to sediment transport and $\Gamma_T$ as the variable free surface of the body of water.
%The bathymetry, $b(X,t)$, and the free surface elevation, $\zeta(X,t)$, of the body of water are used in the parameterization of $\Gamma_B$ and $\Gamma_T$: 
%\begin{subequations}
%\begin{align}
%\Gamma_B &= \{(X,-H_0+b(X,t)):X\in \mathbb R^d\}, \\
%\Gamma_T &= \{(X,\zeta(X,t)):X\in \mathbb R^d\},
%\end{align}
%\end{subequations}
Thus, the domain $D_t$ is defined by $(X,z) \in \mathbb R^d \times \mathbb R$ where $-H_0+b(X,t) < z < \zeta(X,t)$.
%
%
%In the presented work, a dispersive wave hydro-sediment-morphodynamic model developed in \cite{kazhyken_etal_2021_1} is employed as a mathematical model for numerical simulations of hydro-sediment-morphodynamic processes in the Ria Formosa lagoon. 

The dispersive wave hydro-sediment-morphodynamic model is defined over a domain $\Omega \subset \mathbb R^d$ as:
\begin{equation}\label{Eq:GNHSM}
\partial_t \boldsymbol q + \nabla \cdot \boldsymbol F(\boldsymbol q) + \boldsymbol{D}(\boldsymbol q)  = \boldsymbol S(\boldsymbol q),
\end{equation}
where the vector of unknowns $\boldsymbol q$ and the flux matrix $\boldsymbol F(\boldsymbol q)$ are:
\begin{equation}
\boldsymbol q = \begin{Bmatrix} h \\ h \mathbf u \\ hc \\ b \end{Bmatrix}, \quad
\boldsymbol F(\boldsymbol q) = \begin{Bmatrix} h \mathbf u \\ h\mathbf u \otimes \mathbf u + \frac 1 2 g h^2 \mathbf I \\ h c \mathbf u \\ \mathbf q_b \end{Bmatrix},
\end{equation}
the source term $\boldsymbol S(\boldsymbol q)$ is defined:
\begin{equation}
\boldsymbol S(\boldsymbol q) = \begin{Bmatrix}  \frac{E-D}{1-p}  \\ -gh \nabla b - \frac{\rho_s-\rho_w}{2\rho}gh^2 \nabla c-\frac{(\rho_0-\rho)(E-D)}{\rho(1-p)} \mathbf u + \mathbf f \\ E-D \\  - \frac{E-D}{1-p} \end{Bmatrix},
\end{equation}
%
%
%the dispersive term $\boldsymbol{D}(\boldsymbol q)$ is defined,
%
%
%\begin{equation}
%\boldsymbol D(\boldsymbol q) = \begin{Bmatrix} 0 \\ \mathbf w_1 - \alpha^{-1} g h \nabla \zeta \\ 0 \\ 0  %\end{Bmatrix},
%\end{equation}
%
%
where $\mathbf u$ is the water velocity, $h(X,t) = \zeta(X,t) + H_0 - b(X,t)$ is the water depth, $\mathbf f$ comprises additional momentum source terms, $g$ is the  constant of gravitational acceleration, and $\mathbf I \in \mathbb R^{d\times d}$ is the identity matrix. Suspended solids are modeled by a sediment advection process of quantity $hc$, where $c$ is the volume concentration of sediment in the water-sediment mixture, with sediment entrainment, $E$, and deposition, $D$, rates as the source terms defined through the empirical equations from \cite{li_duffy_2011} and \cite{cao_etal_2004}, respectively. The coupled transport of water and sediments also result in additional terms in the momentum and continuity equations, where $p$ is the bed porosity, $\rho_w$ and $\rho_s$ are the water and the sediment densities, $\rho$ and $\rho_0$ are the water-sediment mixture and saturated bed densities defined as $\rho=(1-c)\rho_w+c\rho_s$ and $\rho_0=(1-p)\rho_s+p\rho_w$. In addition to the suspended load, effects of the bed load transport on bed morphology is introduced though the bed load sediment flux $\mathbf q_b$  taken from the empirical Grass model, see \cite{grass_1981}. Finally, we incorporate dispersion into the momentum equation by a variant of the Green-Naghdi equations in the term $\boldsymbol D(\boldsymbol q)$ in~\eqref{Eq:GNHSM}, see Bonneton \emph{et al.}~\cite{bonneton_etal_2011} and our previous work \cite{kazhyken_etal_2021_1} for complete definitions.

\section{Finite Element Mesh Development}
The governing equations are discretized with the algorithm successively developed in \cite{kazhyken_etal_2021_0, kazhyken_etal_2021_1}. The method relies on a Strang operator splitting technique from \cite{strang_1968} such that the model is split into two separate parts: (1) the shallow water hydro-sediment-morphodynamic (SHSM) equations (e.g. see Cao \emph{et al.} \cite{cao_etal_2004}) obtained by dropping the dispersive term of the equations, and (2) the dispersive correction part where the wave dispersion effects on flow velocities are introduced into the model through the dispersive term. The numerical simulation is then propagated in time through a  successive application of numerical solution operators developed for these two parts of the model. In this work the numerical solution operators for the SHSM equations from \cite{kazhyken_etal_2021_1} and for the dispersive part from \cite{samii_and_dawson_2018} are utilized. Both of these operators are discretized using hybridized discontinuous Galerkin methods. The readers are encouraged to consult the original sources in \cite{samii_and_dawson_2018, kazhyken_etal_2021_0,kazhyken_etal_2021_1} for further details.

\begin{figure}
\center
\includegraphics[width=4in]{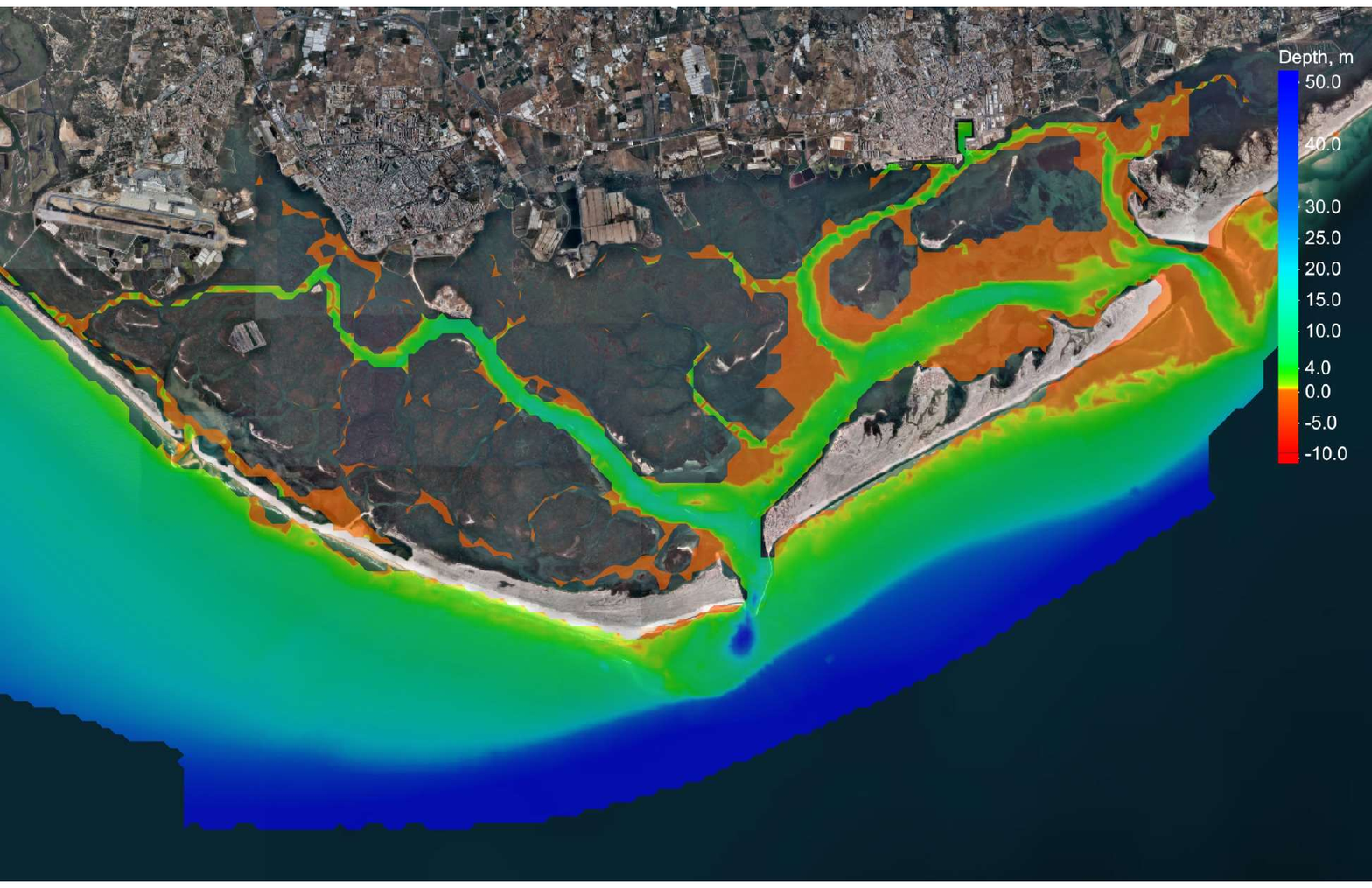}
\caption{Bathymetric data for the Ria Formosa lagoon provided by the Portuguese Hydrographic Institute \cite{HIP}.}
\label{Fig:HIP_bath}
\end{figure}
\begin{figure}
\center
\includegraphics[width=4in]{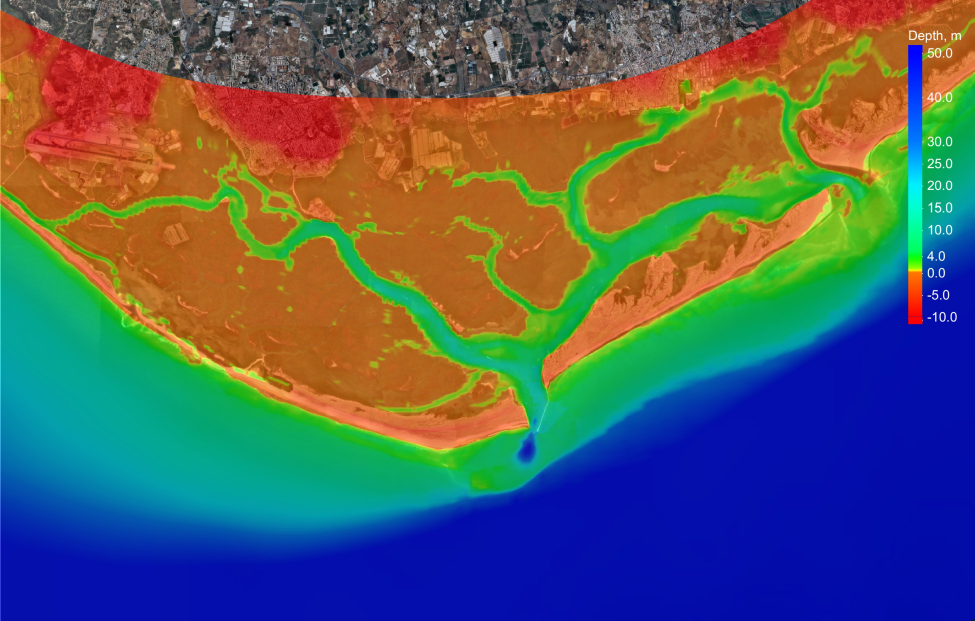}
\caption{Bathymetric data collected for the Ria Formosa lagoon under the SCORE project \cite{SCORE}.}
\label{Fig:SCORE_bath}
\end{figure}

\begin{figure}
\center
\includegraphics[width=4in]{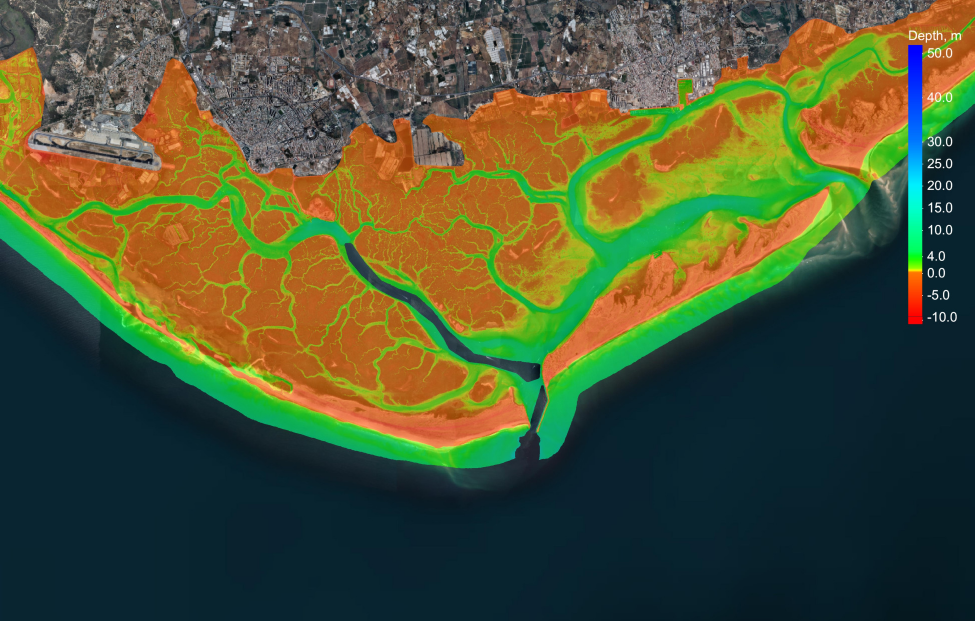}
\caption{LiDAR measurements of the bathymetry in the Ria Formosa lagoon \cite{gorbena_etal_2017, SCORE_DB}.}
\label{Fig:LIDAR_bath}
\end{figure}

\begin{figure}
\centering
\begin{subfigure}{0.45\textwidth}
\includegraphics[width=\textwidth]{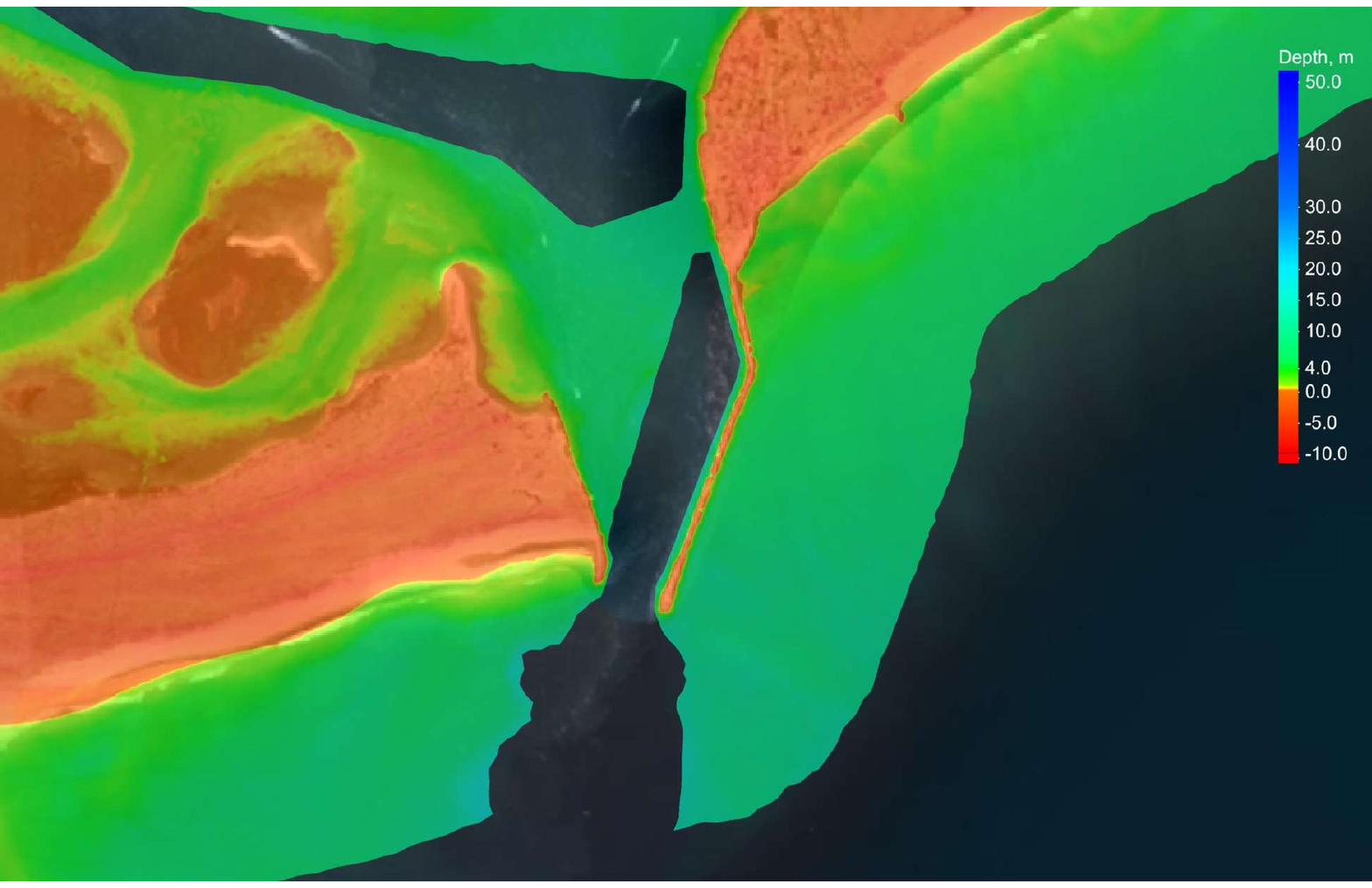}
\caption{LiDAR bathymetric data}
\end{subfigure}
\begin{subfigure}{0.45\textwidth}
\includegraphics[width=\textwidth]{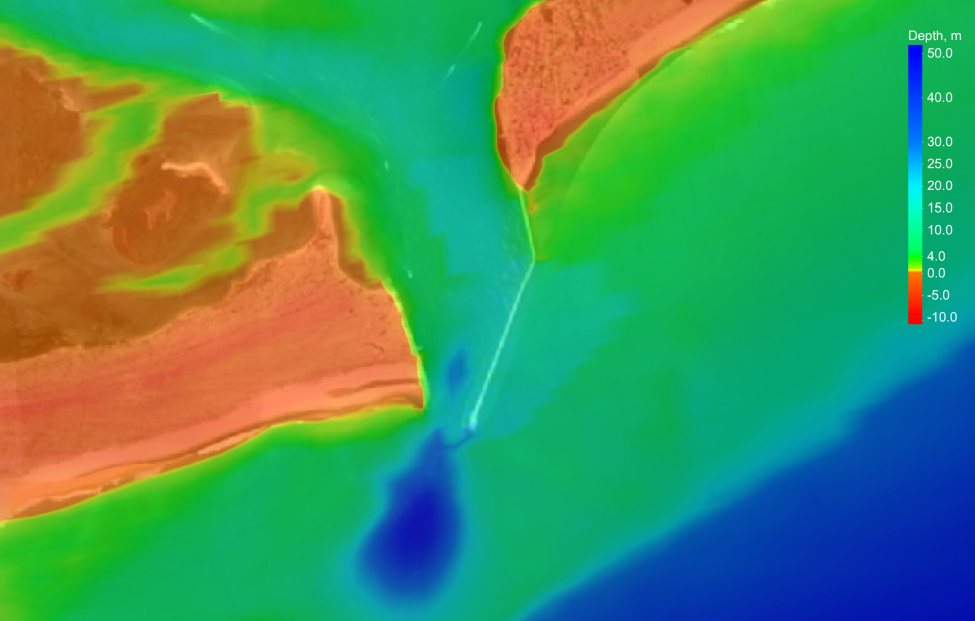}
\caption{SCORE project bathymetric data}
\end{subfigure}
\caption{Comparison of the level of detail in the bathymetic data around the Faro-Olh\~ao inlet.}
\label{Fig:faro_score_lidar}
\end{figure}

As the most hydrodynamically significant, the western circulation cell, serviced by the Anc\~ao, Faro-Olh\~ao and Armona inlets, has been chosen as an application area for the dispersive wave hydro-sediment-morphodynamic model developed in \cite{kazhyken_etal_2021_1}. To this end, a model representation of the cell has been developed. Initially, bathymetric data has been gathered for this area of modeling interest. Four sources for the data has been identified: (1) Portuguese Hydrographic Institute bathymetric model of the Ria Formosa lagoon (cf. Fig. \ref{Fig:HIP_bath}) \cite{HIP}, (2) bathymetric surveys performed under the SCORE project (cf. Fig. \ref{Fig:SCORE_bath}) \cite{SCORE, gorbena_etal_2017}, (3) LiDAR bathymetric data of the coast of Portugal (cf. Fig. \ref{Fig:LIDAR_bath}) \cite{gorbena_etal_2017, SCORE_DB}, and (4) EMODnet Bathymetry Digital Terrain Model (DTM 2018) data \cite{EMODnet}. These data sources have varying levels of detail and coverage, for instance, unlike the LiDAR data, the SCORE project data does not capture the breakwaters of the Faro-Olh\~ao inlet (cf. Fig. \ref{Fig:faro_score_lidar}). The bathymetric surveys in the lagoon completed by the Portuguese Hydrographic Institute have been focused on gathering the data over the lagoon's major channels and in the nearby offshore areas only; therefore, the data does not cover the salt marshes and barrier islands of the lagoon. In contrast, the SCORE project data does provide the bathymetric data over the lagoon's salt marshes and barrier islands. The LiDAR data also provides the data over these areas of the lagoon; however, it lacks data in areas where the water depth has been too large to carry out LiDAR measurements, e.g. inside the channel that connects the city of Faro with the Atlantic Ocean. Finally, EMODnet is the only data source that provides bathymetry measurements in the open ocean; however, since EMODnet aims to provide the global bathymetric data, the data has a low resolution and cannot be used as a source of bathymetric data inside the lagoon. 

These four sources of the bathymetric data have required a degree of preprocessing to generate the bathymetric profile for the western circulation cell and areas adjacent to it. The LiDAR data has been selected as the primary source for the data. The missing data inside the channels has been retrieved from the data provided by the Portuguese Hydrographic Institute. The data for the near shore areas has been lifted from the SCORE project data. Finally, the data for the open ocean has been collected from the EMODnet data source. Therefore, whenever the bathymetry of a point inside the model in development has been requested, the bathymetry would be retrieved, in the descending order of priority, from: (1) the LiDAR data, (2) the Portuguese Hydrographic Institute data, (3) the SCORE project data, (4) the EMODnet data. 

\begin{figure}[h!]
\center
\includegraphics[width=4in]{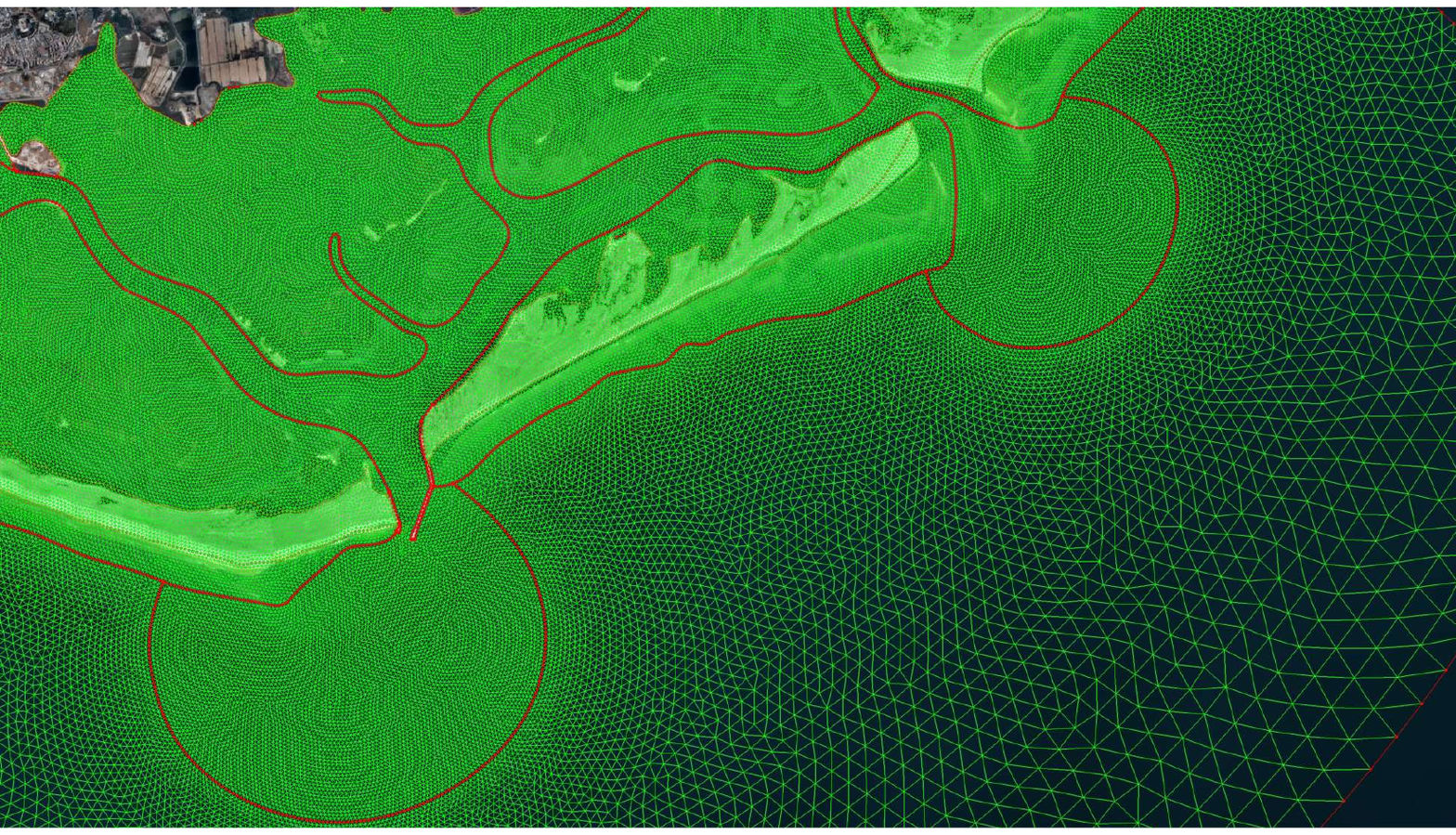}
\caption{Meshing zones defined for the model. Red lines delineate the boundaries between the zones.}
\label{Fig:Zones}
\end{figure}

\begin{figure}[h!]
\center
\includegraphics[width=4in]{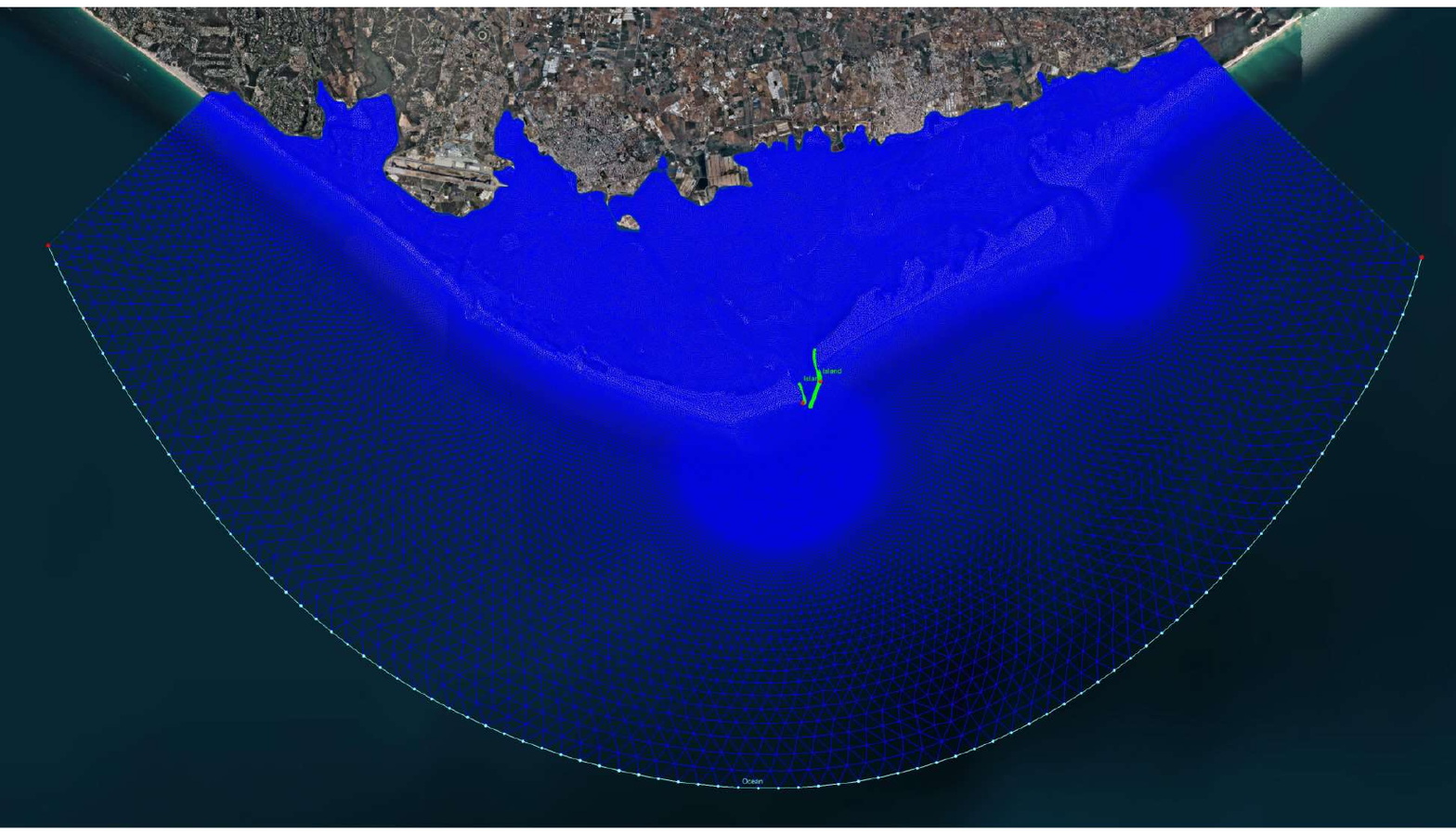}
\caption{Unstructured finite element mesh of the western circulation cell of the Ria Formosa lagoon.}
\label{Fig:Mesh}
\end{figure}

After collecting and preprocessing the bathymetric data, an unstructured finite element mesh for the western circulation cell of the Ria Formosa lagoon has been created with the use of Aquaveo SMS software package. A few observations have been made before the mesh generation process: (1) the majority of the flow parameters variation has been expected to be observed around the tidal inlets of the lagoon; therefore, it has been desired to provide a finer mesh resolution inside and in the vicinity of the inlets; (2) in areas with a higher depth and away from the lagoon inlets a coarser mesh resolution has been requested to conserve computational resources required to run simulations with the developed model; (3) the Aquaveo SMS software package provides a facility to distinguish between land and water elements; therefore, areas around the barrier islands and salt marshes have required a separate meshing treatment. Fig. \ref{Fig:Zones} provides an outline of the meshing zones that have been defined for the model. 
A total of three zone types have been identified: (1) zones around the salt marshes and barrier islands that would transition between wet and dry elements during the high and low tides, respectively; (2) zones inside and in the vicinity of the tidal inlets, and along the lagoon channels where a finer mesh resolution has been desired; (3) zones away from the lagoon inlets in the open ocean that would have a coarser mesh resolution. The final mesh, which has over $1.5 \cdot 10^5$ triangular elements with diameters ranging from 30 to 500 m, is presented in Fig. \ref{Fig:Mesh}, where it is visible that the element density is higher inside the lagoon and in the vicinity of the inlets, and lower in the open ocean areas. 

\begin{figure}[h!]
\center
\includegraphics[width=4in]{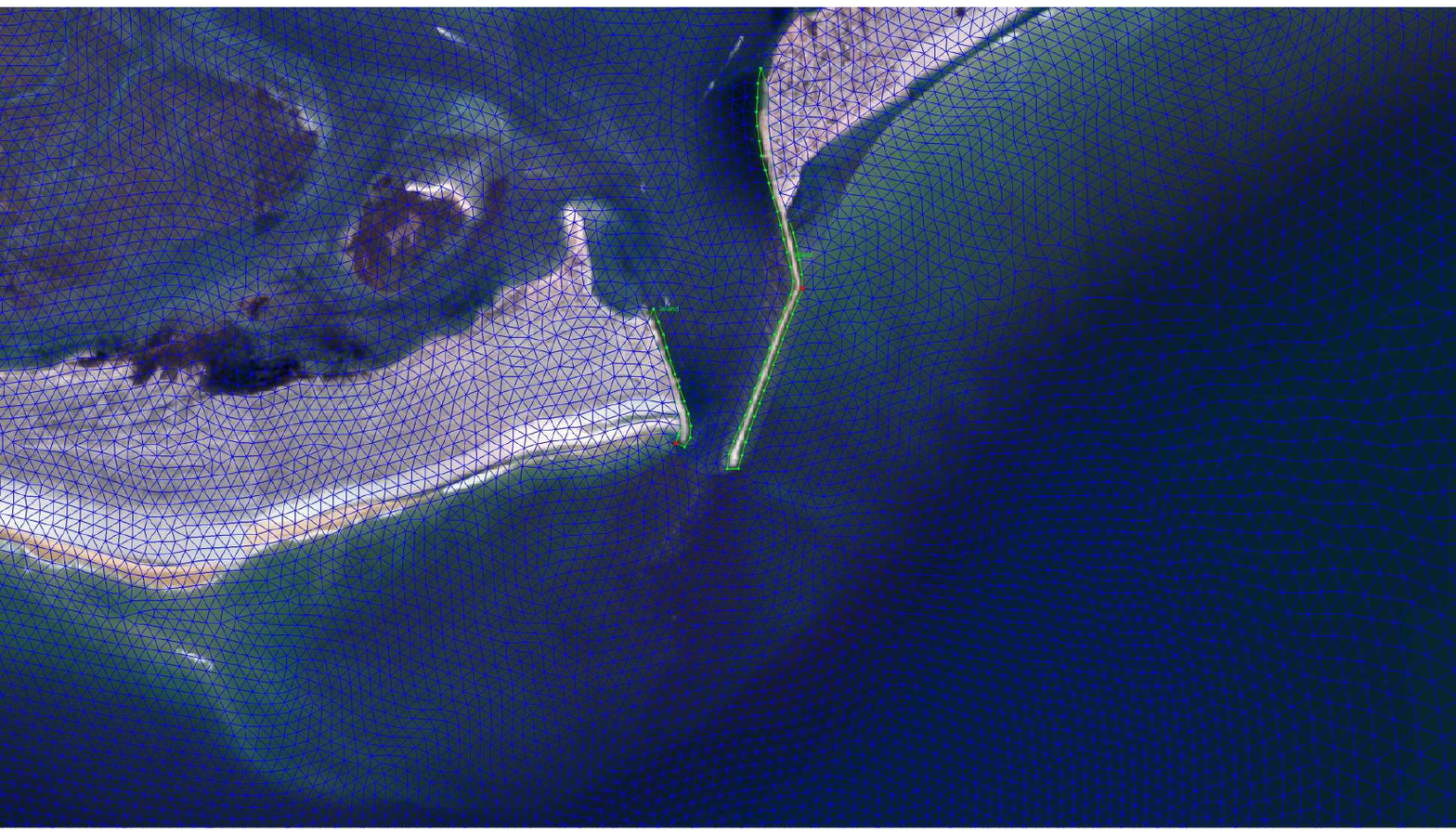}
\caption{Mesh in the vicinity of the Faro-Olh\~ao inlet with two mesh cutouts delineated in green that model the inlet breakwaters.}
\label{Fig:Faro_mesh}
\end{figure}

The northern onshore boundary, and the eastern and western boundaries positioned normal to the shoreline have been modeled as reflective walls, i.e. only flow that is tangent to the wall is permitted, and any incident waves are reflected off the wall. The southern open ocean boundary of the mesh has been used to force the hydrodynamic part of the dispersive wave hydro-sediment morphodynamic model with tidal waves. The forcing has been implemented by prescribing the free surface elevation at the open ocean boundary that is characterized by the tidal constituents (K1, O1, P1, Q1, M2, S2, N2, and K2) obtained through OceanMesh2D software \cite{pringle_etal_2021} and utilizing TPXO (TOPEX/Poseidon satellite) tidal dataset from \cite{egbert_erofeeva_2002}. To the best of our knowledge, prescribing a boundary condition other than a reflective wall for the dispersive correction part of the developed numerical model remains ambiguous. A number of authors suggest using absorbing-generating layers to generate wave conditions inside a problem domain (see e.g. Duran and Marche \cite{duran_and_marche_2017}); however, these layers must have a length of the same magnitude as the wavelength of an imposed wave which is unfeasible for tidal waves that have a typical wavelength on the scale of hundreds of kilometers. Therefore, the open ocean boundary conditions have been imposed through the nonlinear shallow water equations, and the dispersive correction has not been applied in the vicinity of the boundary (5 layers of elements of the mesh away from the open ocean boundary). Finally, in order to model the breakwaters of the Faro-Olh\~ao inlet, two cutouts in the mesh that trace the breakwaters have been done, and the boundaries of the cutouts have been prescribed as reflective walls (cf. Fig. \ref{Fig:Faro_mesh}). 

The Manning's formula has been employed to characterize the bottom friction force in the developed numerical model:
\begin{equation}
\mathbf f = \frac{gn^2}{h^{1/3}}{\vert\mathbf{u}\vert\mathbf{u}},
\end{equation}
with the Manning’s roughness coefficient defined according to the water depth in Table \ref{Tab:Manning}.
\begin{table}[h!]
    \centering
    \begin{tabular}{|c|c|}
\hline
Water depth (m) & Manning's $n$\\
\hline
h $\leq$ -2.5 & 0.032\\
-2.5 $<$ h $\leq$ -2.0 & 0.028\\
-2.0 $<$ h $\leq$ -1.5 & 0.023\\
-1.5 $<$ h $\leq$ -1.0 & 0.020\\
-1.0 $<$ h $\leq$ 0.0 & 0.019\\
0.0 $<$ h $\leq$ 0.5 & 0.024\\
0.5 $<$ h $\leq$ 1.0 & 0.026\\
1.0 $<$ h $\leq$ 3.0 & 0.025\\
3.0 $<$ h $\leq$ 10.0 & 0.023\\
h $>$ 10.0 & 0.022\\
\hline
    \end{tabular}
    \caption{The Manning’s roughness coefficient variation in the Ria Formosa lagoon as reported by Dias \emph{et al.} in \cite{dias_etal_2009}.}
    \label{Tab:Manning}
\end{table}
Sediment entrainment and deposition rates for the suspended load transport have been calculated using the sediment properties from Table \ref{Tab:Sediment}, the sediment density $\rho_s=2650\,\,\text{kg\,m}^{-3}$ \cite{pacheco_etal_2011}, and the bed porosity $p=0.40$.
\begin{table}
    \centering
    \begin{tabular}{|c|c c|}
\hline
Inlet & $d_{50}$ (mm) & $\theta_{c} $\\
\hline
Anc\~ao & 0.88 & 0.029\\
Faro-Olh\~ao & 0.86 & 0.029 \\
Armona & 0.61 & 0.031 \\
\hline
    \end{tabular}
    \caption{Sediment properties at the inlets of the western circulation cell \cite{pacheco_etal_2011}.}
    \label{Tab:Sediment}
\end{table}
The scaling factor for the sediment entrainment model has been chosen as $\phi=3\cdot10^{-4}$. For the bed load transport the Grass model has been used with $A=5\cdot10^{-3}$. The water density has been selected as $\rho=1025\,\,\text{kg\,m}^{-3}$ \cite{carrasco_etal_2018}. To avoid shocks at the open ocean boundary, simulations have been ramped up for 12 hours (43,200 seconds), i.e. the prescribed free surface elevation at the open ocean boundary has been scaled with:
\begin{equation}\label{Eq:ramp}
r = \tanh\frac{2t}{43200},
\end{equation}
where $r$ is the ramping scaling factor, and $t$ is the simulation time in seconds.

\section{Results and Discussion}

\begin{figure}[h!]
\centering
\begin{subfigure}{0.475\linewidth}
\includegraphics[width=2in]{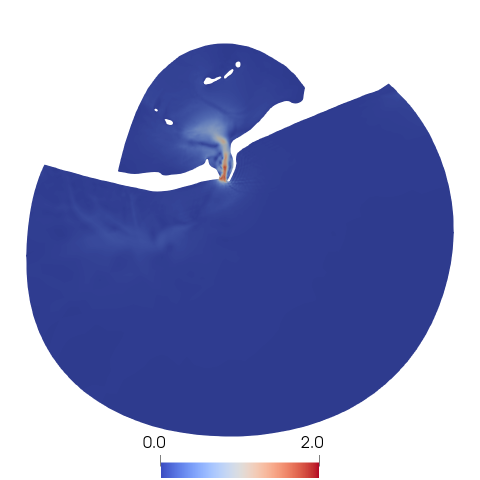}
\caption{$\lvert\textbf{u}\rvert_{\text{GN}}$}
\end{subfigure}
\begin{subfigure}{0.475\linewidth}
\includegraphics[width=2in]{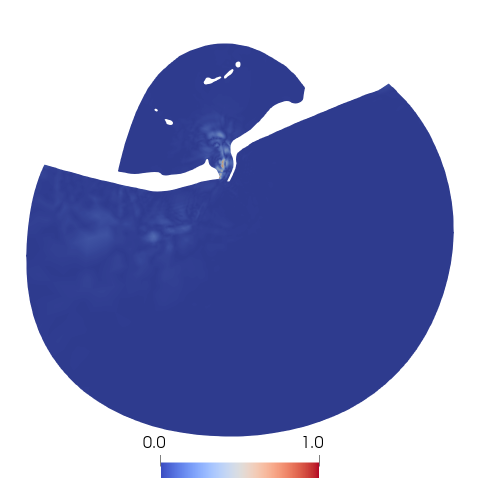}
\caption{$\lvert\lvert\textbf{u}\rvert_{\text{GN}}-\lvert\textbf{u}\rvert_{\text{NSWE}}\rvert$}
\end{subfigure}
\caption{Velocity fields around the time of the peak inflow velocity at the Faro-Olh\~ao inlet. Figure generated by reproducing results from our past work~\cite{kazhyken_etal_2021_0}.}
\label{Fig:maxIN}
\end{figure}

\begin{figure}[h!]
\centering
\begin{subfigure}{0.475\linewidth}
\includegraphics[width=2in]{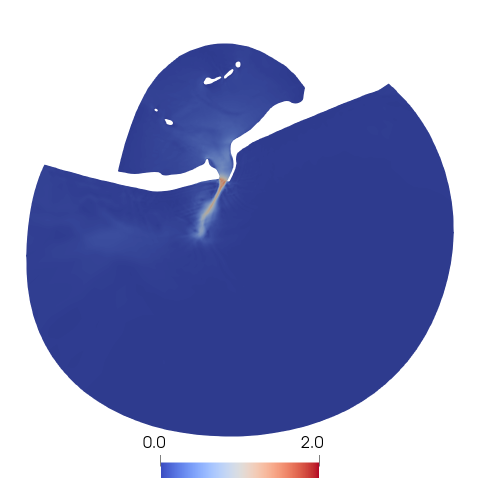}
\caption{$\lvert\textbf{u}\rvert_{\text{GN}}$}
\end{subfigure}
\begin{subfigure}{0.475\linewidth}
\includegraphics[width=2in]{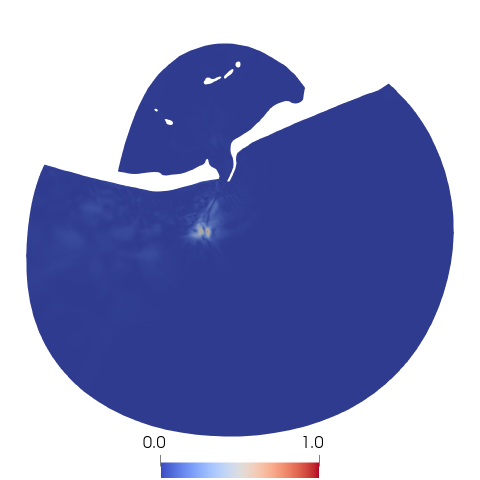}
\caption{$\lvert\lvert\textbf{u}\rvert_{\text{GN}}-\lvert\textbf{u}\rvert_{\text{NSWE}}\rvert$}
\end{subfigure}
\caption{Velocity fields around the time of the peak outflow velocity at the Faro-Olh\~ao inlet. Figure generated by reproducing results from our past work~\cite{kazhyken_etal_2021_0}.}
\label{Fig:maxOUT}
\end{figure}

As a proof of concept that the Green-Naghdi equations could be used to simulate hydrodynamic processes in coastal areas with irregular geometries, an attempt to simulate water waves in the Faro-Olh\~ao inlet has been made in \cite{kazhyken_etal_2021_0}. The water waves have been simulated for 2 days, 12 hours of which have been spent ramping up the simulation with the ramping factor from Eq.(\ref{Eq:ramp}), both with the Green-Naghdi and nonlinear shallow water equations. The velocity profiles around the times of the peak inflow and outflow velocities at the neck of the inlet are presented in Fig.\ref{Fig:maxIN} and Fig.\ref{Fig:maxOUT}, respectively.
The hydrodynamic model has been able to successfully simulate the water waves with the Green-Naghdi equations over the irregular shaped unstructured mesh. The magnitude of difference between the velocity profiles obtained with the Green-Naghdi and nonlinear shallow water equations shows that there is a considerable dissimilarity between these two computations.  These differences are concentrated in regions near the inlet, where the additional wave dynamics are strong. However, as shown in Fig.\ref{Fig:maxIN} and Fig.\ref{Fig:maxOUT}, in most of the computational domain, the differences are negligible. Since the nonlinear shallow water equations are embedded in the Green-Naghdi equations, the dissimilarities can be attributed to actions of the dispersive term of the Green-Naghdi equations that captures additional water wave dynamics through terms that are $O(\mu^2)$ consistent with the incompressible Euler equations, and not present in the nonlinear shallow water equations. Although these terms have the capacity to capture dispersive wave effects, the dissimilarities may not be attributed to the dispersive effects as the water motion in this simulation has been forced with an M2 tidal wave. This example shows that using the Green-Naghdi equations may contribute additional water wave dynamics into the presented hydro-sediment-morphodynamic model through the dispersive term.

\begin{figure}[h!]
\centering
\includegraphics[width=4in]{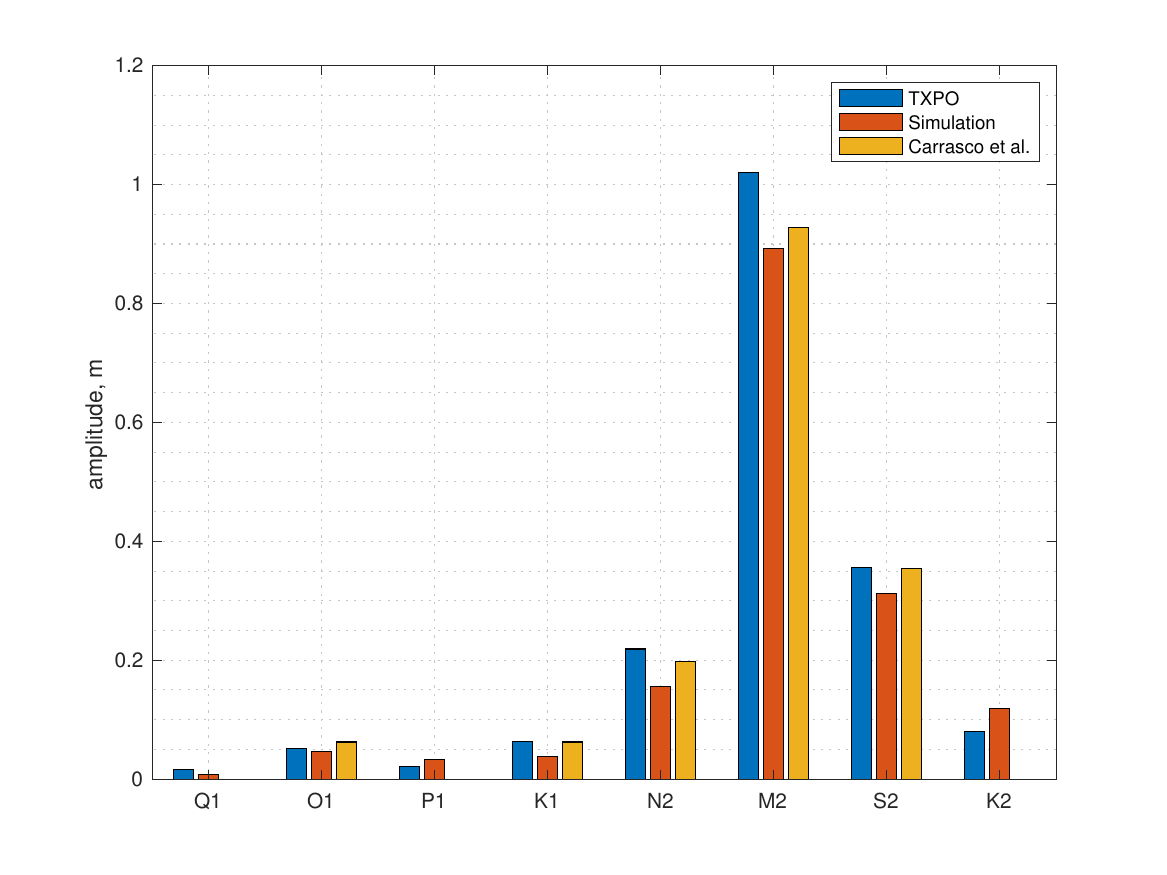}
\caption{Amplitudes of the tidal spectrum at St.1 compared with the results from TXPO in \cite{egbert_erofeeva_2002} and Carrasco \emph{et al.} in \cite{carrasco_etal_2018}.}
\label{Fig:S1}
\end{figure}

\begin{figure}[h!]
\centering
\includegraphics[width=4in]{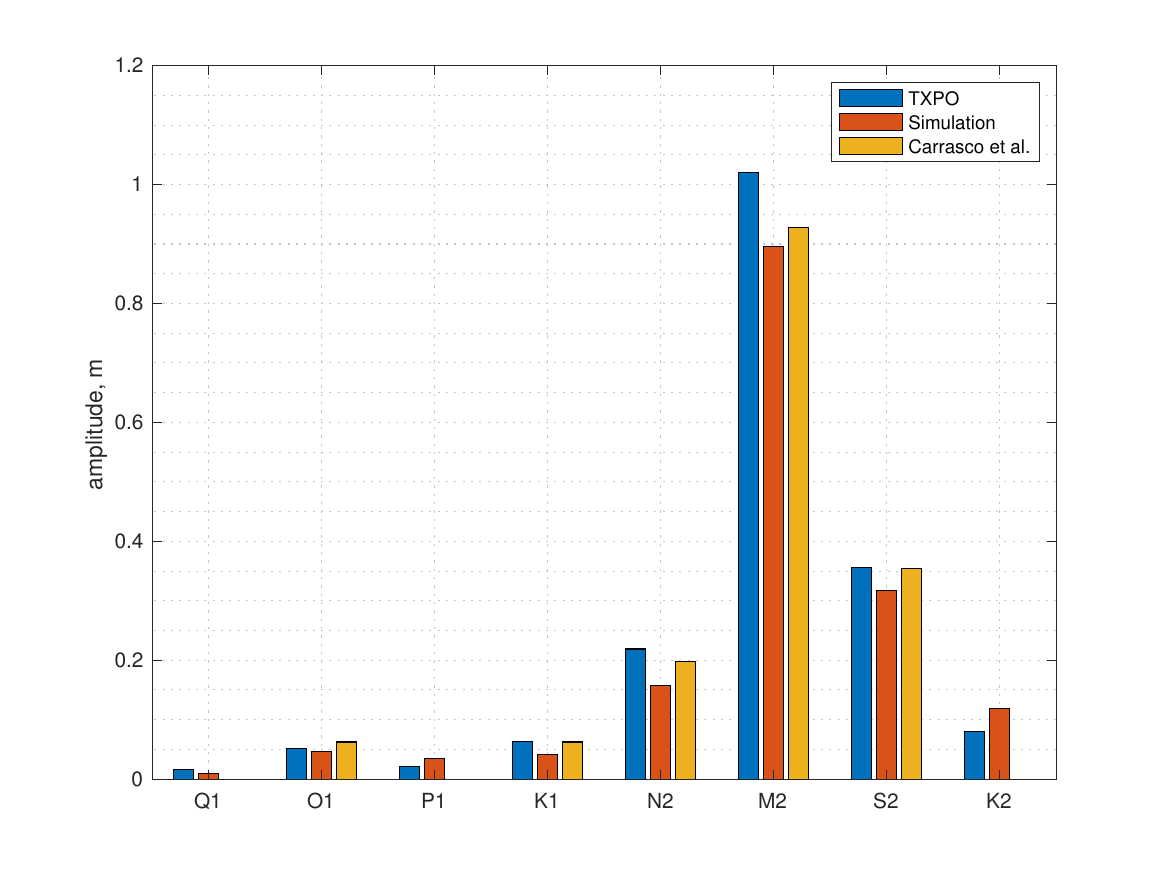}
\caption{Amplitudes of the tidal spectrum at St.2 compared with the results from TXPO in \cite{egbert_erofeeva_2002} and Carrasco \emph{et al.} in \cite{carrasco_etal_2018}.} 
\label{Fig:S2}
\end{figure}

\begin{figure}[h!]
\centering
\includegraphics[width=4in]{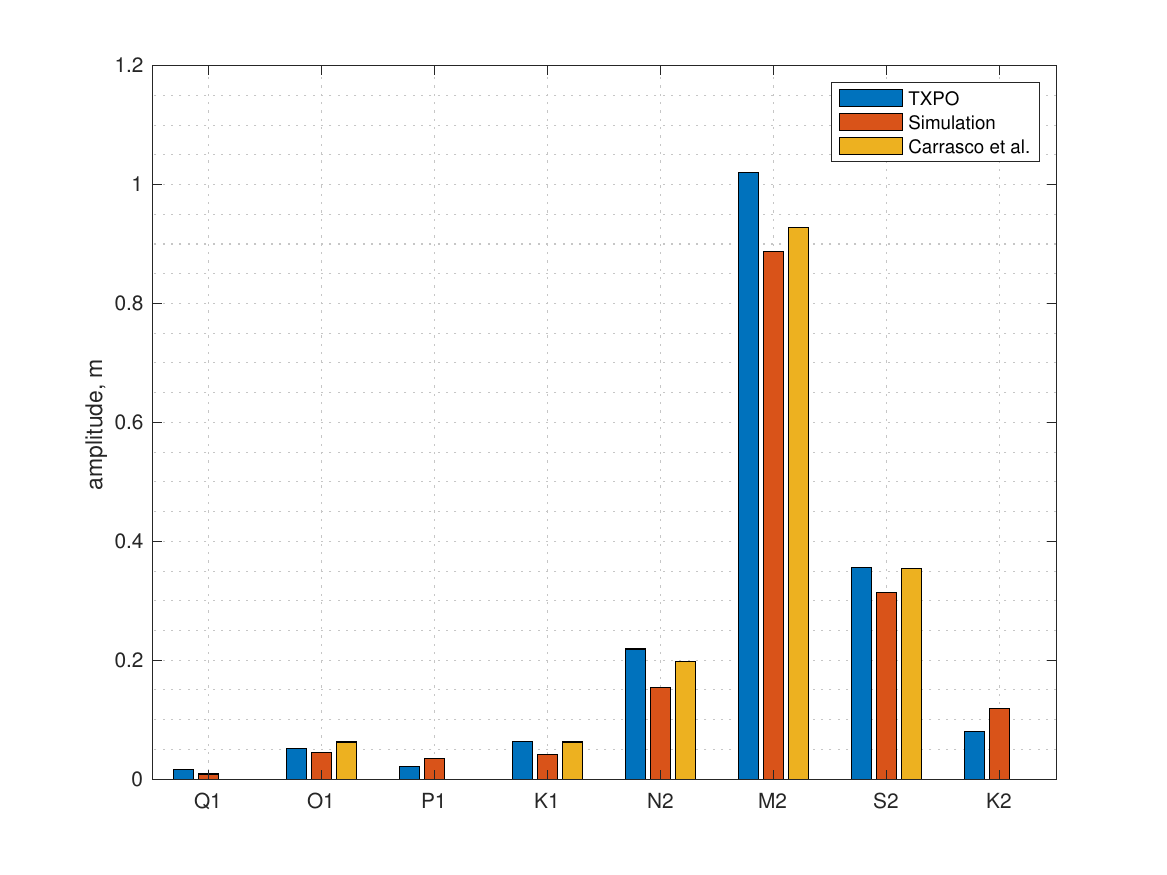}
\caption{Amplitudes of the tidal spectrum at St.3 compared with the results from TXPO in \cite{egbert_erofeeva_2002} and Carrasco \emph{et al.} in \cite{carrasco_etal_2018}.}
\label{Fig:S3}
\end{figure}

To validate the developed numerical finite element model of the western circulation cell, it has been decided to compare the tidal spectrum generated with the data collected while running simulations with the model to the spectra from TXPO in \cite{egbert_erofeeva_2002} and Carrasco \emph{et al.} in \cite{carrasco_etal_2018}. To this end, three points that correspond to three water level measuring stations St.1 (-7.8685, 36.9740), St.2 (-7.8705, 36.9661), and St.3 (-7.9178, 37.0025) have been selected for the analysis. Simulations for the analysis have been ran while disabling sediment transport and dispersive correction parts of the model, i.e. only the hydrodynamic part in the nonlinear shallow water equations has been used for the simulations. After running the simulations for 56 days, the water level data collected at the three measuring stations has been used to perform the analysis. The amplitudes of the tidal spectrum for 8 major constituents Q1, O1, P1, K1, N2, M2, S2, and K2 at these measuring stations are presented in Fig.\ref{Fig:S1}-\ref{Fig:S3}. 
The calculated amplitudes are compared to the spectra from  TXPO in \cite{egbert_erofeeva_2002} and Carrasco \emph{et al.} in \cite{carrasco_etal_2018}. It is worth noting that \cite{carrasco_etal_2018} does not provide the tidal amplitude data for Q1, P1, and K2 constituents, thus the comparison omits these data points.
From these figures we notice that the M2 constituent has the largest discrepancy with respect to the TXPO model. It is important to note that the TXPO model (as indicated in Section 5 of \cite{egbert_erofeeva_2002}) is a global model and thus may not have sufficient resolution in local areas. Furthermore, the bathymetry data used in both TXPO and in \cite{carrasco_etal_2018} likely differs slightly from ours as the lagoon is highly dynamic with large changes in sediment deposition and scouring. When comparing to the results in \cite{carrasco_etal_2018} we also note that the models cover different portions of the offshore domain and thus apply boundary forcing in different manners which may influence the results. Another source that may lead to the observed discrepancies is the application of sea bottom friction, while we employ the values listed in Table 1, \cite{carrasco_etal_2018} uses a more limited range of only three Manning's n values and the TXPO model uses a different quadratic bottom friction formulation. Nonetheless, it is evident that the model has the capacity to  predict tidal waves in the lagoon with reasonable accuracy when compared to the other model results from literature.

In the next stage the full model has been used to simulate hydro-sediment-morphodynamic processes in the western circulation cell. The dispersive term of the developed numerical model requires a solution of a set of global equations, which in this particular simulation example has nearly $10^6$ degrees of freedom; therefore, running a simulation of this magnitude requires a utilization of a parallel computing cluster. To this end, a total of 1536 cores (32 Intel Xeon Platinum 8160 ("Skylake") nodes 48 cores per node each at the Stampede2 supercomputer at The Texas Advanced Computing Center) have been employed to run 2 day simulations, 12 initial hours of which have been used to ramp the simulations up with the ramping factor from Eq.(\ref{Eq:ramp}), which took 6 hours of wall-clock time to complete. We select the 2 day simulation period to observe the effect of a few tidal cycles; the M2 tidal period is approximately 12 hours, so the period covers three full cycles after the initial ramping period and represents an appropriate period presentation of the developed model. While appropriate for the current demonstration, 2 days may be insufficient to capture longer term sediment transport effects. Finally, we point out that the high computational cost means that the model run for the GN case required about 50 hours continuous running on the employed supercomputer.

\begin{figure}[h!]
\center
\includegraphics[width=4in]{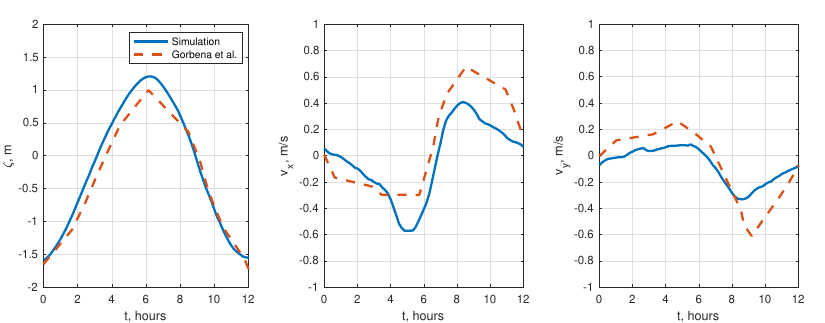}
\caption{Flow parameters at the ADCP measurement site compared with the measurements reported by Gonz\'alez-Gorbe\~na \emph{et al.} in \cite{gorbena_etal_2017}.}
\label{Fig:Gorbena}
\end{figure}

\begin{figure}[h!]
\center
\includegraphics[width=4in]{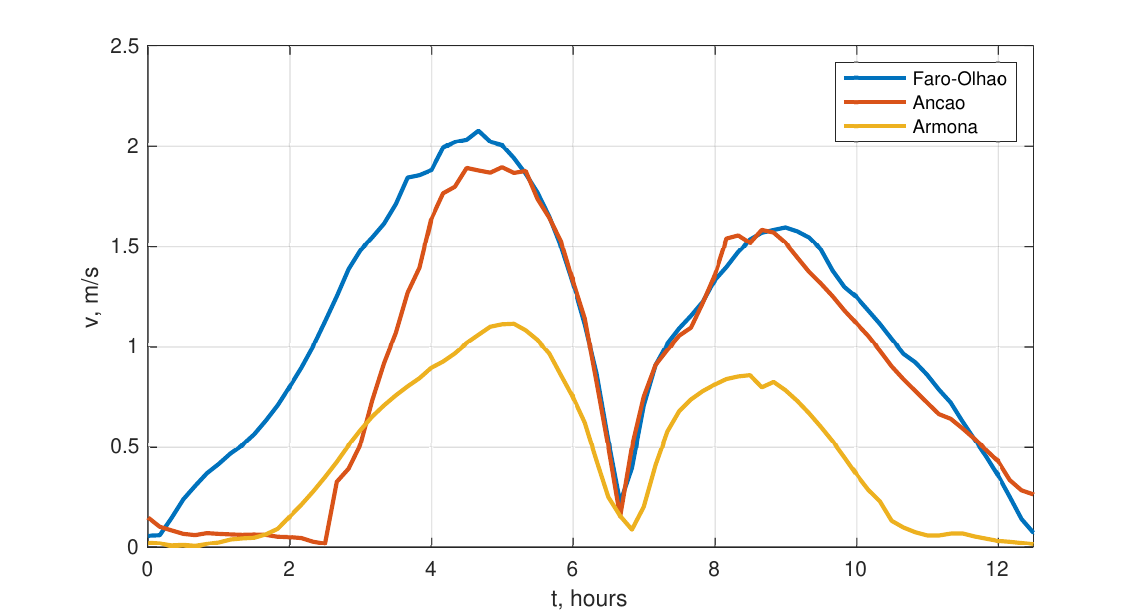}
\caption{Velocity magnitudes at the western circulation cell inlets during the M2 tidal wave cycle.}\label{Fig:Velocity}
\end{figure}

The hydrodynamic results of the simulation have been compared with the acoustic Doppler current profile (ADCP) measurements performed and reported by Gonz\'alez-Gorbe\~na \emph{et al.} in \cite{gorbena_etal_2017}. The measurements have been collected in the vicinity of the Faro-Olh\~ao inlet just north of the inlet inside the Faro channel. The results are presented in Fig.~\ref{Fig:Gorbena}; and it is evident that the simulation results are in a satisfactory agreement with the measurements, in particular, the free surface elevation results are in a good agreement with the measurements.
Fig.~\ref{Fig:Velocity} presents velocity magnitudes at the inlets during the full cycle of the M2 tidal wave. It is worth noting, that the velocity magnitudes at the inlets are similar to the maximum velocities reported by Salles \emph{et al.} in \cite{salles_etal_2005} for Anc\~ao and Armona inlets at 1.35 and 1.05 $\text{ms}^{-1}$, respectively, and reported by Pacheco \emph{et al.} in \cite{pacheco_etal_2008} for the Faro-Olh\~ao inlet at 2.2 $\text{ms}^{-1}$. 

\begin{figure}
\center
\includegraphics[trim=0in 0in 0in 0in, clip=true, width=4in]{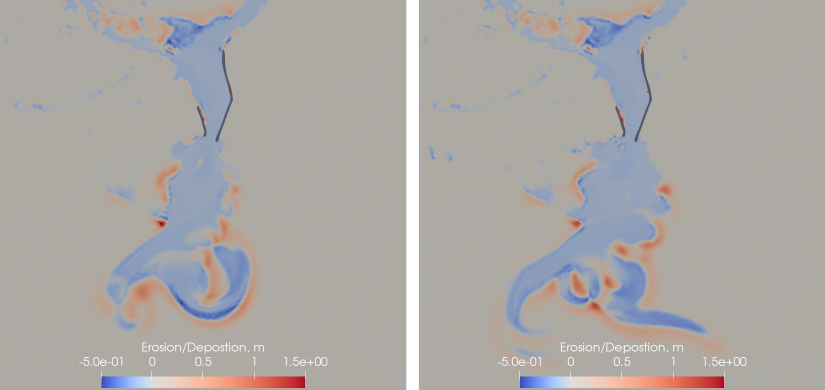}
\caption{Net erosion and deposition results for the Faro-Olh\~ao inlet after 2 day simulations with the shallow water (left) dispersive wave (right) hydro-sediment-morphodynamic models.}\label{Fig:net_sediment}
\end{figure}

In the developed model for the western circulation cell, the Anc\~ao inlet is shallow and dries up during the low tide; this fact explains the lack of flow at the inlet during hours 0 to 2. Although the velocity magnitudes in the Anc\~ao inlet are comparable to the Faro-Olh\~ao inlet, due to the shallowness and narrowness of the Anc\~ao inlet, the total discharge through that inlet is relatively small compared to the much deeper and wider Faro-Olh\~ao inlet. Finally, the velocity magnitudes at the natural Armona inlet are similar to the values presented by Dias \emph{et al.}  (see Figure 6 in \cite{dias_etal_2009}), with peak values between 0.6 and 0.7 $\text{ms}^{-1}$. While the overall trend and velocity magnitudes match well with past published model results, a one-to-one comparison is not trivial as the aforementioned model results were performed and published more than a decade ago. The Ria Formosa area and its inlets are highly dynamic and constantly changing thereby making comparisons beyond overall trends and magnitudes not feasible. As observed in the water wave simulation in the Faro-Olh\~ao inlet, the velocity distribution around the inlet varies between the simulations completed with the Green-Naghdi and nonlinear shallow water equations. Therefore, a simulation using the shallow water hydro-sediment morphodynamic model has been run for the western circulation cell. Since the sediment entrainment and deposition rates, as well as the bed load flux values are dependent of the flow velocity, it is expected that the net erosion and deposition results will vary between the simulations. The net erosion and deposition results obtained after 2 day simulations with the dispersive wave and shallow water hydro-sediment-morphodynamic models are presented in Fig.~\ref{Fig:net_sediment}. It is evident that the velocity distribution differences, in particular, the way eddies are formed in the ebb flow jet of the Faro-Olh\~ao inlet, have led to the variations in the net erosion and deposition values.

Further inspection of Fig.~\ref{Fig:net_sediment} shows that the two models agree very well inside the inlet as the deposition and erosion patterns are very similar. This area is also noteworthy as quite significant erosion and deposition is observed after the two day simulation period. This indicates that the initial bathymetry and sediment set-up is far from a state of equilibrium further underlining the highly dynamic nature of this region.  
However, outside the inlet we observe different patterns. As shown in Fig.~\ref{Fig:maxOUT}(b), there is a significant difference in the velocity fields of the two models during outflow phases. Since the sediment transport are highly dependent on these fields, we expect a corresponding significant difference in the sediment transport.

The dispersive term of the developed hydro-sediment-morphodynamic model introduces terms into the hydrodynamic part of the model that are $O(\mu^2)$ consistent with the incompressible Euler equations. If these terms have a significant influence on the flow parameters, e.g. if wave dispersion effects are prevalent in water wave regimes, then using the developed model will provide a facility to capture effects the dispersive term has both on hydrodynamics, as well as sediment transport and bed morphodynamics. While the developed dispersive model is able to reproduce physical events, it it significantly more costly that its shallow water counterpart. In terms of computational intensity the inclusion of the dispersive terms leads to an approximate increase in runtime that is ten fold.

\section{Conclusion}

In this paper, we have presented the application of our dispersive hydro-sediment-morphodynamic model from  \cite{kazhyken_etal_2021_1} to the western portion of the Ria Formosa Lagoon. To this end, we have developed a new unstructured finite element mesh covering the region with spatially varying descriptions of bathymetry and Manning's $n$ coefficients. The resulting numerical model is forced with tidal data at the offshore boundary of the domain corresponding to eight principal tidal constituents.  The shallow water hydrodynamic portion of the model is validated by comparison of the tidal spectrum to the TPXO tidal model and the results presented in \cite{carrasco_etal_2018}. Whereas the full Green-Naghdi dispersive model is compared to published gauge data in \cite{gorbena_etal_2017,dias_etal_2009}.

We subsequently use the validated models to simulate the sediment erosion and deposition in the western part of the Ria Formosa Lagoon. The simulations are performed for both the shallow water and the dispersive wave  hydro-sediment-morphodynamic models. Comparison reveals that there are pronounced effects on the sediment transport of incorporating dispersive effects. To the best of our knowledge, this is the first attempt that has been made to use a hybridized discontinuous Galerkin dispersive wave hydrodynamic model, the Green-Naghdi equations, to model not only the hydrodynamic processes, but also the sediment transport and bed morphodynamic processes in a coastal area with a complex irregular geometry.

Future potential directions of work include validation of the developed models through field measurements of sediment and extension of the models to other geographic regions. While the current considered flows are not greatly impacted by meteorological forcing, future work will include wind surface loads in the hydrodynamics, as well as coupling to spectral wave models such as Simulating Waves Nearshore (SWAN) \cite{ris1999third} to take into accounts from short-long wave interactions. While we have demonstrated the model capabilities over a 2 days simulation time, a longer term study covering an extended period should be performed and optimally validated against other model and/or collected data.

\section*{Acknowledgements}
This work has been supported by funding from the National Science Foundation Grant 1854986, and the Portuguese government through Funda\c{c}\~ao para a Ci\^encia e a Tecnologia (FCT), I.P., under the project DGCOAST (UTAP-EXPL/MAT/0017/2017). Authors would like to acknowledge the support of the Texas Advanced Computing Center through the allocation TG-DMS080016N used in the parallel computations of this work. 

\section*{Data availability}
The datasets generated and analysed during the current study are available from the corresponding author on reasonable request.

\section*{Declarations}
The authors have no competing interests to declare that are relevant to the content of this article.

\bibliography{sn-bibliography}

\end{document}